\documentclass[16pt]{IEEEtran}
\usepackage{algorithm}
\usepackage{algpseudocode}
\usepackage{booktabs} % 提供更美观的表格
\usepackage{graphicx} % 如果需要插入图片等内容
\usepackage{cite}

\usepackage{caption}
\usepackage[english]{babel}
\usepackage{anyfontsize} % 允许使用任意字体大小
\renewcommand\normalsize{\fontsize{14}{16}\selectfont}
\usepackage[letterpaper,top=2cm,bottom=2cm,left=3cm,right=3cm,marginparwidth=1.75cm]{geometry}

\usepackage{amsmath}
\usepackage{graphicx}
\usepackage[colorlinks=true, allcolors=blue]{hyperref}

\title{Heat: Satellite's meat is GPU's poison}
\author{Zhehu Yuan, Jinyang Liu, Guanqun Song, Ting Zhu}

\begin{document}
\maketitle

\begin{abstract}
In satellite applications, managing thermal conditions is a significant challenge due to the extreme fluctuations in temperature during orbital cycles. One of the solutions is to heat the satellite when it is not exposed to sunlight, which could protect the satellites from extremely low temperatures. However, heat dissipation is necessary for Graphics Processing Units (GPUs) to operate properly and efficiently. In this way, this paper investigates the use of GPU as a means of passive heating in low-earth orbit (LEO) satellites. Our approach uses GPUs to generate heat during the eclipse phase of satellite orbits, substituting traditional heating systems, while the GPUs are also cooled down during this process. The results highlight the potential advantages and limitations of this method, including the cost implications, operational restrictions, and the technical complexity involved. Also, this paper explores the thermal behavior of GPUs under different computational loads, specifically focusing on execution-dominated and FLOP-dominated workloads. Moreover, this paper discusses future directions for improving GPU-based heating solutions, including further cost analysis, system optimization, and practical testing in real satellite missions.

\end{abstract}

\section{Introduction}
Thermal management is one of the most critical challenges for satellite operations in space, especially for small satellites that face extreme temperature variations between the sunlit and eclipse phases of their orbits. For satellites, temperature fluctuations can reach 300 degrees Celsius in a few hours \cite{colozza2006evaluation}. Such a large temperature difference in a short period of time could cause serious results, including thermal fatigue \cite{mirza2017cyclic, kaewkham2022mechanical}, disable of electric parts \cite{patterson2009evaluation}, accelerate battery wear and tear \cite{low_temp_anodes, low_temp_electrolytes, low_temp_leo_batteries, nasa_battery_rnd}, etc, which will finally reduce the life span of satellites. 

One of the solutions is using passive thermal control systems \cite{yildirim2023passive, kumar2022passive}, including but not limited on thermal straps, passive radiator, and heat pipes. These methods could passively maintain the temperature of satellites within a suitable range. However, these designs are delicate and vulnerable\cite{luojia_passive_thermal, deep_space_habitat_thermal}. If any part of the system are broken, the whole system may be failed. Also, these systems cannot be changed after the satellite is launched \cite{active_vs_passive_2024}. If something unexpected happens outside the design, people have no way to modify the system accordingly. In this way, the passive thermal control systems are not reliable. Moreover, their thermal control ability is relatively weak. For those extreme thermal control requirements, passive thermal control systems have nothing to do with them \cite{passive_thermal_control_2023, passive_cubesat_2019, passive_design_applied_2022}.

To mitigate these limitations of passive thermal control systems, active thermal control methods are designed \cite{li2016ehd, liang2015fluid, autonomous2023, themis2024, smartskin2021}, such as heaters and coolers. These devices could actively control the temperature of the satellites, which make them robust against partial failures and flexible to be adapted for out-of-design situations. When some of them are disabled, the others could increase their power to produce more heat. When out-of-design situations happen, people can manually control the thermal controller to adapt to the situations. However, their prices present significant challenges, particularly for smaller satellites with limited budget \cite{nasa_small_spacecraft}. In addition, these active temperature control devices require energy to operate, leading to inefficiencies and energy waste.

Heat is the meat of satellites, but it is also the poison of GPU. During the computation of GPU, a large amount of waste heat are generate. If heat dissipation is not managed promptly, the GPU's efficiency will decline, and it may even become inoperative \cite{price2014optimizing, data_center_temperature, gpu_reliability}.

In this study, we propose a novel approach to thermal control of satellites that utilizes GPUs as a heater in satellites. In one way, satellites are benefited with the efficiency of GPU in producing heat. On the other hand, the power consumed on heating is not wasted, while they are also used to do computation. Even better, while heating the satellites, GPUs are also cooled down, which helps them to maintain a better efficiency.

In this paper, we made the following contributions:

\begin{itemize}
    \item We proposed a thermal control system that using GPU as a heater.
    \item We compared the prices, sizes, and power of existed representative GPUs and satellite heaters.
    \item We proposed a way to control the heat produced by GPU through job assignment.
    \item We measured the heat produced by GPU running different jobs.
\end{itemize}

\section{Background}

\subsection{GPU Side}

%Energy Produce

%https://gamersnexus.net/hwreviews/2830-nvidia-gtx-1080-ti-fe-review-and-game-benchmarks
When GPUs are running, they produce a significant amount of energy in the form of heat because almost all the electrical energy they consume is converted into heat. For example, the NVIDIA GTX 1080 Ti, a high-end model released in 2017, has a Thermal Design Power (TDP) of 250 watts. This means that it can consume up to 250 watts of electrical power under full load, and almost all of this power is dissipated as heat\cite{nvidia_2017}.

%Temperature Benefit

%https://arxiv.org/pdf/1407.8116

Temperature plays a critical role in optimizing GPU power efficiency. The subthreshold leakage current, which is a significant contributor to power consumption, is proportional to \( T^2 e^{-b/T} \), where \( T \) is the temperature. This relationship shows that when the temperature increases, the leakage current increases, leading to higher power consumption. In contrast, maintaining lower temperatures reduces leakage current and improves GPU power efficiency. Therefore, lowering the temperature is essential to achieve optimal performance per watt in GPUs\cite{price_optimizing_2015}.

\subsection{Satellite Side}

%Heat Required and Power Saving

%https://ieeexplore.ieee.org/document/9993478

%https://doaj.org/article/70a52cd104e74cdbb6233b23af06b565

%https://research.tudelft.nl/en/publications/the-effect-of-melting-point-and-combination-of-phase-change-mater

Satellites experience cyclic extreme thermal conditions due to the alternating sunlit and eclipse zones, significantly affecting operational efficiency. For example, deployable solar panels (DSPs) are particularly susceptible to temperature fluctuations caused by the space environment. Maintaining the operating temperature range of solar cells (\(-40^{\circ}C\) to \(+125^{\circ}C\)) is crucial, as inadequate thermal management can lead to reduced efficiency or functional failure\cite{budiantoro_thermal_2022}. 

Small satellites face unique thermal challenges due to their small heat capacity, compact designs, limited radiator areas, and power constraints, making thermal control even more complex than for larger satellites \cite{elshaer_thermal_2023}. For example, a small satellite with an internal heat load of approximately 9.2 W may require thermal systems that account for variations of up to 13 W to effectively manage uncertainties \cite{elshaer_thermal_2023}.

To address these challenges, thermal control methods are created to ensure that satellite components operate within their designated temperature ranges in an extreme environment. An approach is a passive thermal control system (TCS), as demonstrated by Budiantoro et al. (2022), which uses surface coatings such as black anodized and SG121FD white paint on DSP to regulate heat absorption and radiation\cite{budiantoro_thermal_2022}. This method is power efficient and cost effective, but is entirely based on material properties, making it less adaptable to rapidly changing thermal conditions or extreme temperature fluctuations, which may impact system performance. Another passive method involves the use of heat pipes for localized cooling, as explored by Elshaer\cite{elshaer_effect_2023}. Heat pipes efficiently transfer excess heat to radiative surfaces without consuming power during operation, addressing high power heat dissipation challenges\cite{elshaer_effect_2023}. However, they are limited in that they cannot actively generate heat, making them unsuitable for addressing extreme cold conditions. Additionally, a hybrid approach combines passive radiative coatings with active heaters to dynamically manage fluctuating temperatures\cite{elshaer_thermal_2023}. This hybrid system balances power efficiency by relying on passive elements in moderate conditions and activating heaters only when needed, making it highly adaptable but introducing additional complexity, higher development costs, and occasional power consumption. Together, these methods offer varied solutions tailored to different satellite designs and operational requirements, with trade-offs in flexibility, efficiency, and resource demands.

%Cooling Constrain

%https://www.nasa.gov/smallsat-institute/sst-soa/thermal-control/

\section{Our Design: GPU Heater}

\begin{figure}[htbp]
    \captionsetup{font={normalsize, rm}}
    \centering
    \includegraphics[width=1\linewidth]{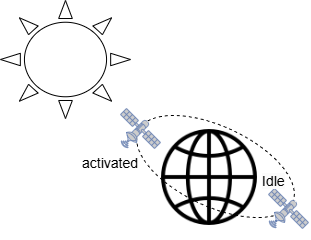}
    \caption{The two stages of GPU based on the location of the satellite}
    \label{fig:two_stage}
\end{figure}

As shown in figure \ref{fig:two_stage}, our design follows a two-stage process based on the local of the satellite:
\begin{enumerate}
    \item \textbf{Activation During Eclipse:} When the satellite is in the eclipse phase, GPUs are activated to perform computational tasks. The heat generated during these operations substitutes for conventional heaters, maintaining the required temperature for sensitive components.

    \item \textbf{Deactivation in Sunlit Phases:} During sunlit periods, the satellites are heated by sun light, and the temperature dramatically increases, so the heaters are not required. Instead, the high temperature becomes harmful for the operation of GPUs. As a result, the best choice for GPUs are remaining idle.
\end{enumerate}

\subsection{Advantages of GPU-Based Heating}
GPU-based heating possesses multiple advantages. To begin with, they are cheaper than traditional heating systems. In addition, operating GPUs under low temperature conditions, as in space, reduces the subthreshold leakage current and improves power efficiency, as described in \cite{price_optimizing_2015}. This phenomenon ensures better performance-per-watt ratios during computations, and we will further discuss it in section \ref{sec: w&t}.

Moreover, using traditional satellite heaters, the energy is only used for heating. However, with GPUs as heaters, heat is only the side product, while running GPUs itself can still fulfill the requirement of satellite computation \cite{bousquet2000onboard, airbus2019payload, george2017hybrid, ott2020gpu, ott2023ai}. Such satellite computing not only shifts computation from the ground to space, but also optimizes data transfer both among satellites and between satellites and ground stations. This, in turn, reduces power consumption and alleviates bandwidth limitations.

\subsection{Challenges and Solutions}

Using GPUs as heater of satellites, we cannot simply replace the traditional satellite heaters with GPUs. Instead, we still face some challenges.

First, unlike traditional satellite heaters, we cannot directly control the amount of heat produced by GPUs. A straight forward solution is to shut down GPU when we don't need it, and turn it on when we need it. However, this solution cannot precisely control the heat production. For example, if we aim to slightly reduce heat production rather than completely eliminate it, this straightforward approach is ineffective. To achieve precise control over heat production, we can regulate the tasks performed by the GPU. When we need more heat, we will assign high heat production jobs to GPU. When we need less hearing, we will assign low-heat production jobs to GPU. We will discuss the detailed relationship between jobs and heat production in Section \ref{sec: w&t}.

\begin{figure}[htbp]
    \centering
    \includegraphics[width=1\linewidth]{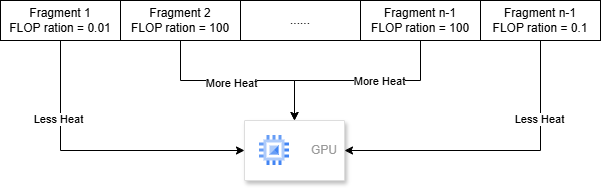}
    \caption{The two stages of GPU based on the location of the satellite}
    \label{fig: segmentation}
\end{figure}

\begin{figure}[htbp]
    \centering
    \includegraphics[width=1\linewidth]{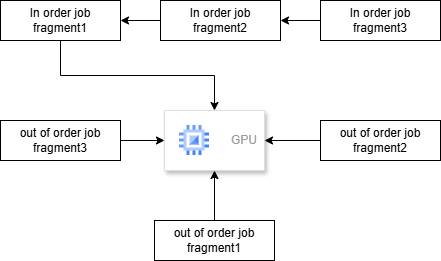}
    \caption{The two stages of GPU based on the location of the satellite}
    \label{fig: order}
\end{figure}

Moreover, unlike running GPUs on ground, GPUs cannot keep running one job for a long time in satellites. In Sunlit Phases, the GPUs need to remain idle. When we want to control heat production, we may force GPUs to switch jobs. To avoid losing all progress when switching jobs or shutting down GPUs, we will divide the jobs into several fragments, each with the same running time. As shown in figure \ref{fig: segmentation}, after running each fragment, the job progress will be saved. In this way, we only lose the progress of a job fragment when a job is switched or the GPUs are shut down. For each job fragment, we will also estimate its heat production effect based on the ratio between FLOPs and memory access. In this way, we could assign proper job fragments to GPUs to achieve precise heat control. As shown in figure \ref{fig: order}, for those in order jobs, where the process of one segment is dependent on the results from previous segment, only the first unfinished segment in order can be picked and assigned to GPU. For those out of order jobs, any segments can be assigned to GPU in any order.

\section{Experiment Setup}

In this experiment, we investigated the thermal behavior of GPUs in the context of satellite applications. The test platform consisted of an RTX 3070 laptop \cite{nvidia_rtx3070_laptop}, which was used to simulate typical satellite conditions. The primary objective was to examine the performance under two distinct computational scenarios: execution-dominated and FLOP-dominated. These scenarios were represented by two separate kernels: kernel1 and kernel2, each designed to highlight different characteristics of GPU usage in terms of computational load and memory access patterns. Each of the kernels are run 50 times with $testSize = 300000$ and $workload = 150$.

\subsection{Execution-Dominated Workload}
An \textbf{execution-dominated} workload (Algorithm \ref{execution dominated}) is one in which the number of floating-point operations (FLOPs) per memory access is high, specifically more than 100 FLOPs per memory access. In this scenario, the workload performs many computational operations relative to the number of memory accesses. The GPU's compute units are highly engaged, leading to high throughput in terms of FLOPs. 

This approach is chosen because The GPU is highly utilized for computational tasks, resulting in a more intensive execution of arithmetic operations. Furthermore, high computational intensity leads to greater power consumption and therefore more heat generation, which can affect the GPU temperature. In addition, understanding execution-dominated workloads helps evaluate the effect of high computational workloads on the GPU's temperature and performance. It is particularly useful for designing systems where computation is the primary task, such as complex simulations.

\begin{algorithm}
\caption{Execution-Dominated kernel}
\label{execution dominated}
\begin{algorithmic}[1]
\State Initialize $i$ as the thread index based on blockIdx and threadIdx
\State Load value from array $a$ at index $i$ into variable $b$
\State Initialize $d \gets 0$
\For{$j \gets 0$ \textbf{to} $testSize$}
    \State Perform the following floating-point operations on $b$
    \State $b \gets b \times a[i]$
    \State $b \gets b \times 3.0$
    \State $b \gets b / 6.0$
    \State $b \gets b / 2.0$
    \State $b \gets b + 1$
\EndFor
\State Store the final value of $b$ into array $c$ at index $i$
\end{algorithmic}
\end{algorithm}

\subsection{Memory-Dominated Workload}
A \textbf{Memory-dominated} workload (Algorithm \ref{memory dominated}) is one where the number of floating-point operations per memory access is low, specifically less than 0.01 FLOP per memory access. In this scenario, the workload involves frequent memory accesses but fewer computational operations per memory access. This type of workload is more memory bound, meaning that the system spends more time accessing and transferring data than performing actual computations. It is chosen because the workload involves more memory read and Additionally, studying FLOP-dominated workloads helps us understand the impact of memory-bound tasks on GPU performance and temperature. This is useful for optimizing heat dissipation in memory-intensive applications where computation is not the bottleneck.

Since memory dominated jobs complete much faster than execution dominated jobs, we have to do more such execution to make them have similar time consumption. To do this, for each round, we run the memory dominated jobs $workload$ extra times (line 3).

\begin{algorithm}
\caption{Memory-Dominated Kernel}
\label{memory dominated}
\begin{algorithmic}[1]
\State Initialize $i$ as the thread index based on blockIdx and threadIdx
\State Initialize $b \gets 0$
\For{$j \gets 0$ \textbf{to} $testSize \times workload$}
    \State Access memory at index $i+j$ and update $b$
    \State $b \gets b + a[i+j]$
    \State $b \gets b + a[i+j+1]$
    \State $b \gets b + a[i+j+2]$
    \State $b \gets b + a[i+j+3]$
    \State $b \gets b + 1$
\EndFor
\State Store the final value of $b$ into array $c$ at index $i$
\end{algorithmic}
\end{algorithm}

\section{Results}

\subsection{The Relationship between Workload and Temperature}
\label{sec: w&t}

Due to technology limit, we cannot remove the cooler of GPU, which will automatically active to cool down GPU when the temperature of GPU is too high. However, we could still observe the effect of different workload on temperate. The results of the experiment show the behavior of the GPU under both execution-dominated and memory-dominated conditions in terms of temperature and computational time.

\begin{figure*}[htbp]
    \centering
    \includegraphics[width=0.49\textwidth]{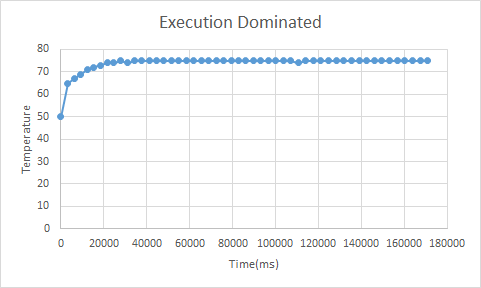}
    \includegraphics[width=0.49\textwidth]{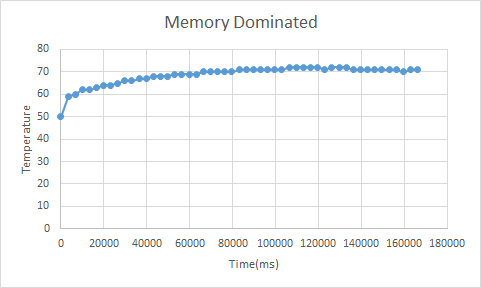 }
    \caption{The temperature of GPU against time}
    \label{fig:overall}
\end{figure*}

From figure \ref{fig:overall}, we can observe that the GPU temperature increases under both workloads. As the GPU's cooler activates, the rate of temperature increase slows. However, even with the cooler in operation, the GPU temperature for execution-dominated workloads rises significantly faster than that for memory-dominated workloads, ultimately reaching a higher peak. Overall, this indicates that FLOP operations have a greater impact on GPU temperature compared to memory access operations.

\begin{figure*}[htbp]
    \centering
    \includegraphics[width=0.49\textwidth]{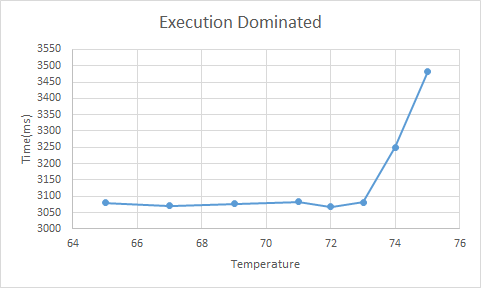}
    \includegraphics[width=0.49\textwidth]{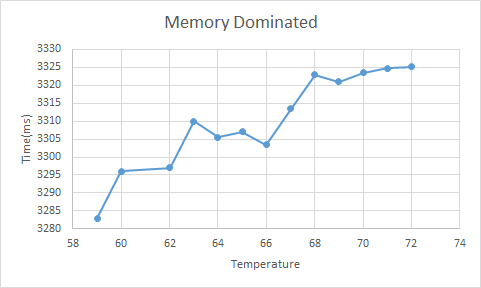}
    \caption{The running time of each round of execution under different GPU temperature}
    \label{fig:single}
\end{figure*}

In figure \ref{fig:single}, we test the running time of each round of execution for both workload under different GPU temperature. For the execution-dominated workload, the running time remains constant until the GPU temperature reaches 73°C, after which it increases dramatically for about 200 ms per degree. For the memory-dominated workload, the running time keeps growing as the the GPU temperature increases. But such increase in running time is relatively mild, for 3.24 ms per degree on average.

As a conclusion, compared to the memory-dominated workloads, the execution-dominated workload has larger effect on GPU temperature and is more sensitive to GPU temperature. Thus, we recommend to run execution-dominated jobs when the temperature is low and run memory-dominated jobs when the temperature is high. In this way, the execution-dominated jobs could be benefited with low temperature and benefit the satellite for producing more heat, while the memory-dominated jobs are benefited with insensitive for GPU temperature and  benefit the satellite for producing less heat.

\subsection{Power, Size, and Price}

\begin{table*}[htbp]
\centering
\caption{Product Comparison Table}
 \resizebox{\textwidth}{!}{
\begin{tabular}{lccc}
\toprule
\textbf{Product}                          & \textbf{Price(\$)} & \textbf{Power(W)} & \textbf{Size(cm\(^3\))} \\ \midrule
MSI GTX 980 GAMING 4G                     & 389                & 165               & 1406.16                  \\
MSI GTX 950 GAMING 2G                     & 299                & 90                & 1368.63                  \\
maxsun RX 550 4GB                  & 84                 & 35                & 731.675                  \\
ASRock Intel ARC A380 6GB                  & 110                 & 75                & 1062.936                  \\
Omega Polyimide Heater Kit                & 62.89              & 5                 & 0.164                    \\
Minco Polyimide Thermofoil        & 106                & 7.5               & 0.098                    \\ \bottomrule
\end{tabular}
}
\label{tab:product_comparison}
\end{table*}

In this section, we compared the powers, sizes, and prices of representative GPUs\footnote{Price data are collected from www.amazon.com}\footnote{Powers and size data are collected from www.maxsun.com, www.msi.com, and www.intel.com} and traditional satellite heaters\footnote{Data collected from www.omega.com and www.minco.com}. 

From table \ref{tab:product_comparison}, we could observe that the most price-efficient GPU is ASRock Intel ARC A380 6GB consuming 1.47 dollars/Watt, and the most size efficient GPU is MSI GTX 980 GAMING 4G consuming 8.5$cm^3 / Watt$. The most price efficient traditional satellite heater is Omega Polyimide Heater Kit consuming 12.58 dollars/Watt, while the most size efficient traditional satellite heater is Minco Polyimide Thermofoil Heaters consuming 0.013$cm^3 / Watt$. In this way, we can conclude that GPUs are significantly more cost-efficient than traditional satellite heaters. For the same level of heat production, the most cost-efficient GPU requires only $1/9$ of the cost of the most cost-efficient traditional satellite heater.

\begin{figure}[htbp]
    \centering
    \includegraphics[width=1\linewidth]{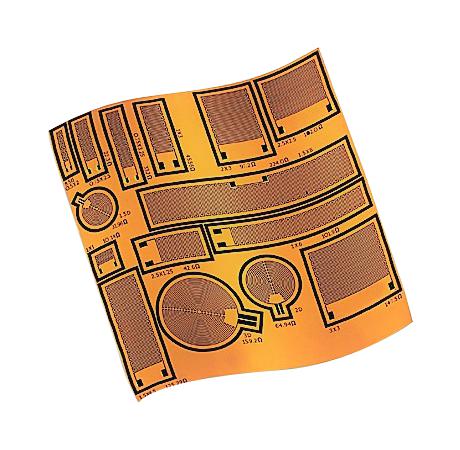}
    \caption{The small size of traditional satellite heaters comes from their extremely thin thickness}
    \label{fig:thin}
\end{figure}

Although GPUs are seriously size-inefficient compared to traditional satellite heaters in theory, they are not so bad in reality. The small size of traditional satellite heaters comes from their extremely thin thickness, as shown in figure \ref{fig:thin}. However, multiple heaters cannot be directly stuck together, and there must be sufficient space between them to facilitate the flow of heat. In this way, significantly more space than theory are required to stuck up the traditional satellite heaters to achieve equivalent power as GPUs.

\section{Future Work}

The current report outlines several areas for future exploration and potential improvements in the field of GPU thermal management for satellites.

\begin{itemize}
    \item \textbf{What to Do When There is No Job for GPU:} Research is needed to explore alternative ways to utilize GPU power when it is not actively involved in computational tasks, or when there is no proper task for it. One of the solutions is to solve long-last puzzles or NP-hard problems, including but not limited on hash collision, Graph Edit Distance, etc. 
    \item \textbf{Heating Experiments on Real Satellites:} Since we do not have real satellites, we can not accurately define the heat requirement of satellites. In the further, we will test our design on real satellites. 
    \item \textbf{Heat Transfer from GPU to Other Parts of the Satellites:} GPUs can produce heat, but such heat are concentrated around the GPUs. Tranditionally, the heat produced by GPUs are spread with air and wind. However, this is not feasible in the space. In the future, we will do more research on Materials Science and Design Science to find a way to transfer the heat produced by GPU to other parts of the satellites.
    \item \textbf{Protection Against Space Environment:} In the space, the environment is different the earth, and many challenges many encountered. For example, the vacuum environment may increase the requirement of compression resistance of the GPU, and we may need to substitute the plastic parts on GPU with stronger material. Also, the space radiation many disrupt the GPU. Additional protection may be necessary, which could lead to increased demands on cost, weight, and size.
    \item \textbf{Job Fragmentation Strategy:} In our design, the jobs should be divided into fragments. If the jobs fragments could be independent from each other, our design could achieve the best performance by freely assign jobs to GPU out of order and only based on the heat requirements. In the further, we will fragment the representative jobs computed on satellites, so that there are little dependency between fragments.
    \item \textbf{Job-Heat Estimation Function} In this study, we found that execution jobs produce more heat than memory access jobs. However, the GPU we have is built in the laptop along with the cooler, and the cooler cannot be disabled. Therefore, we cannot accurately define how much heat will be produced by each kind of jobs. In the future, we could buy a independent GPU, put it into a cooling liquid, and measure the amount of heat produced by measuring the temperature change of the cooling liquid.
\end{itemize}

The rapid advancements in IoT \cite{wire1,wire3,10017581, 9523755,9340574,10.1145/3387514.3405861,9141221,9120764,10.1145/3356250.3360046,8737525,8694952,10.1145/3274783.3274846,10.1145/3210240.3210346,8486349,8117550,8057109,https://doi.org/10.1155/2017/5156164,10189210,MILLER2022100245,DBLP:journals/corr/abs-2112-15169,YAO2020100087,MILLER2020100089,8556650,10.1145/3127502.3127518,10.1145/3132479.3132480,GAO201718}, secure communication \cite{wire2, 10125074,285483,10.1145/3395351.3399367}, and artificial intelligence technologies \cite{10.1145/3460120.3484766,9709070,9444204,ning2021benchmarkingmachinelearningfast,8832180,8556807,8422243,chandrasekaran2022computervisionbasedparking,iqbal2021machinelearningartificialintelligence,pan2020endogenous} have opened new opportunities for optimizing satellite systems, including GPU thermal management. By integrating AI models for workload scheduling, leveraging secure communication protocols to enhance data integrity, and utilizing wireless communication technologies for real-time monitoring, future research can address the unique challenges faced in space environments. These interdisciplinary approaches provide a brighter future for innovative solutions in GPU utilization and heat dissipation strategies.

\section{Conclusion}
In conclusion, this study demonstrates that utilizing GPUs for heating in satellite systems offers a promising alternative to traditional heating methods, particularly in terms of reducing prices and enhancing the overall power efficiency of satellites. The ability of GPUs to generate heat during computational tasks provides a dual benefit, addressing both thermal and computational needs. 

Beside on simply setup GPUs on satellites, we designed a job assignment method to achieve precise control on heat production. Also, to avoid losing all progress on switching jobs or shutting down GPU and to better control the heat production, we fragmented the jobs into several fragments, each with estimated effect on heat production.

Future work should focus on optimizing this approach, exploring alternative power utilization strategies, and conducting real-world experiments in satellite environments to fully assess the feasibility and performance of GPU-based heating systems.

\bibliographystyle{ieeetr}

\bibliography{energy_produce,temperature_benefits,heat_required_Budiantoro,heat_required_Elshaer_1,heat_required_Elshaer_2,thermal_weight,sample,zhu}
\end{document}